\documentclass[runningheads,a4paper]{llncs}

\usepackage[latin1]{inputenc}
\usepackage{amssymb,amsmath}

\usepackage{algorithm,algorithmic}

\usepackage{url}
\urldef{\mailsa}\path|abdoul.ciss@ucad.edu.sn, |
\urldef{\mailsb}\path|sowdjibab@yahoo.fr|

\newcommand{\keywords}[1]{\par\addvspace\baselineskip
\noindent\keywordname\enspace\ignorespaces#1}

\begin{document}

\title{Pairings on Generalized Huff Curves}
\titlerunning{Pairings On Generalized Huff Curves}
\author{Abdoul Aziz Ciss and Djiby Sow}
\authorrunning{Abdoul Aziz Ciss, Djiby Sow}
\institute{Laboratoire d'Algèbre, Codage, Cryptologie, Algèbre et Applications
\\Université Cheikh Anta Diop de Dakar, Sénégal \\
BP: 5005, Dakar Fann\\
\mailsa \mailsb}

\maketitle

\begin{abstract}
This paper presents the Tate pairing computation on generalized Huff curves proposed by Wu and Feng in \cite{Wu}. In fact, we extend the results of the Tate pairing computation on the standard Huff elliptic curves done previously by Joye, Tibouchi and Vergnaud in \cite{Joux}. We show that the addition step of the Miller loop can be performed in $1\mathbf{M}+(k+15)\mathbf{m}+2\mathbf{c}$ and the doubling one in $1\mathbf{M} + 1\mathbf{S} + (k + 12) \mathbf{m} + 5\mathbf{s} + 2\mathbf{c}$ on the generalized Huff curve.
.
\end{abstract}

\section{Introduction}
Pairing computations on elliptic curves were introduced in 1948 by  Weil \cite{Weil}, but their utilization in cryptography is actually quite recent.
In 1993, Menezes, Okamoto and Vanstone in \cite{MOV} used the Weil pairing  to convert the discrete logarithm on some elliptic curves to a discrete logarithm in some extension of the base field.\\
More recenttly, Frey and Rück \cite{Frey}  extends the results of Menezes \emph{et al.} on an even wider category of elliptic curves but with the Tate pairing instead of the Weil pairing.\\
Sakai, Ohgishi and Kasahara \cite{Sakai} and Joux \cite{Joux}  proposed independently in 2000 two cryptosystems using pairings on elliptic curves. In fact, Joux presented a Diffie-Hellman look alike protocol, except it allows three entities (instead of two) to create and exchange a secret key. Sakai, Ohgishi and Kasahara studied the use of pairings in identity-based cryptosystems. The idea of ID-based cryptography was introduced  in 1984 by Shamir \cite{Shamir}. It consist in cryptosystem where the the public key of each entity is directly linked to its identity, which removes the need for its certification by a trusted certification authority.\\
Boneh and Franklin in \cite{Boneh} and Cocks \cite{Cocks} proposed separately in 2001 two identity based encryption schemes, the first one use the Weil pairing, the second one uses properties of quadratic residues. Since then, cryptographic pairings and their applications in cryptosystems have caught numerous researchers attention, and new ID-based protocols have been presented frequently \cite{Boldyreva,Boneh2,Castellucia,Cha,Riyami}.\\
Since cryptographic pairings were gaining more and more importance, many researchers lead studies to families of curves where pairings are efficiently computable \cite{Barreto2,Dupont,Edwards,Galbraith1,Miyaji}, as well as studies on efficient algorithms to compute pairings \cite{Barreto1,Galbraith1}.\\
In \cite{Joye}, Joye, Tibouchi and Vergnaud present efficient formul\ae for computing the Tate pairing on Huff curves. Our contribution in this paper is to extend their results on generalized Huff curves proposed by Wu and Feng in \cite{Wu}. In fact, we show that the Tate pairing on generalized Huff curves is as efficient as in ordinary Huff curves and is a good choice to implement ID-based cryptography.\\
The rest of the paper is organized as follows : in the next section,we recall some basic definition and notation on generalized Huff curves and the Tate pairing. In section 3, we give the main result of the paper, i.e. formul\ae for computing the Tate pairing on generalized Huff curves.

\section{Preliminaries}
\subsection{Generalized Huff curves}
In \cite{Wu}, Wu and Feng extend the Huff elliptic curves by introducing the new form
$$\tilde{\mathcal{H}}_{a,b} : x(ay^2 - 1) = y(bx^2 -1),$$
where  $ab(a - b) \neq 0$.
This new model contains the ordinary Huff curves as particular case.\\
If $a = \mu^2$ and $b = \nu^2$ are squares in  $\mathbb{F}$, pose $x' = \nu x$ and $y' = \mu y$. Therefore, $\mu x'(y'^{2} - 1) =  \nu y'(x'^2 - 1)$.\\
That means all curves of the form $ax(y^2 - 1) = by(x^2 - 1)$ are included in the family of curves $x(ay^2 - 1) = y(bx^2 -1)$, where $a$ and $b$ are quadratic residues in  $\mathbb{F}$. Note that $\tilde{\mathcal{H}}_{a,b}$ is smooth if $ab(a - b) \neq 0$.

\begin{theorem}
Let $\mathbb{F}$ be a field of characteristic different from 2, let $a$ and $b$ be two elements of $\mathbb{F}$, with $a\neq b$. Then, the curve
\begin{equation*}
\tilde{\mathcal{H}}_{a,b} : X(aY^2 - Z^2) = Y(bX^2 - Z^2)
\end{equation*}
is isomorphic over $\mathbb{F}$ to the elliptic curve given by the Weierstrass equation
\begin{equation*}
V^2 W = U(U + aW)(U + bW)
\end{equation*}
via the transformations  $\varphi (X,Y,Z) = (U, V, W)$, where $U = bX - aY$, $V = (b - a)Z$ and  $W = Y- X$. The inverse application is given by $\psi (U, V, W) = (X, Y, Z)$, with $X = U + aW$, $Y = U + bW$ et $Z = V$.\\
\end{theorem}
In affine coordinates, the Huff curve $ x(ay^2 - 1) = y(bx^2 -1)$ defined over  $\mathbb{F}$ is isomorphic to the elliptic curve$y^2 = x(x + a)(x + b)$ over $\mathbb{F}$.\\~~\\
\textbf{Expression of the group law over $x(ay^2 - 1) = y(bx^2 - 1)$ : }\\ Let  $y = y_1 + \lambda(x - x_1) = \lambda x + \mu$ be the equation of the line through $P_1$, $P_2$ $\in\tilde{\mathcal{H}}_{a,b}(\mathbb{F}) $, where  $\lambda$ is the slope of the line through $P_1$ and $P_2$. By the equation of the curve, we obtain  $x(a(\lambda x + \mu)^2 - 1) = (\lambda x + \mu)(bx^2 - 1)$. Let  $S = P_1 + P_2 = (x_3, y_3)$. Then,
$$P_1 + P_2 = \left( \frac{(x_1 + x_2)(ay_1 y_2 + 1)}{(bx_1 x_2 + 1)(ay_1 y_2 - 1)}, \frac{(y_1 + y_2)(bx_1 x_2 + 1 )}{(bx_1 x_2 -1)(ay_1 y_2 + 1)} \right) .$$
Consider now the two points $P_1$ and $P_2$ in projective coordinates, ie.  $P_1 = (X_1, Y_1, Z_1)$ and $P_2 = (X_2, Y_2, Z_2)$, and $U = O = (0, 0, 1)$ as the neutral element of the group law. Let $S = P_1 + P_2 = (X_3, Y_3, Z_3)$. Then,
\begin{flushleft}
$\left\{
   \begin{array}{ll}
    X_{3}=\big(X_{1}Z_{2}+X_{2}Z_{1}\big)\big(aY_{1}Y_{2} + Z_{1}Z_{2}\big)^{2}\big(Z_{1}Z_{2}- bX_{1}X_{2}\big),  \\
    Y_{3}=\big(Y_{1}Z_{2}+Y_{2}Z_{1}\big)\big(bX_{1}X_{2} + Z_{1}Z_{2} \big)^{2}\big(Z_{1}Z_{2}- aY_{1}Y_{2}\big),  \\
  Z_{3}=(b^{2}X^{2}_{1}X^{2}_{2} - Z^{2}_{1}Z^{2}_{2})(a^{2}Y^{2}_{1}Y^{2}_{2} - Z^{2}_{1}Z^{2}_{2}).
   \end{array}
 \right.$
\end{flushleft}

Let $\mathbf{m}$, $\mathbf{s}$ and $\mathbf{c}$ be respectively the costs of the multiplication, squaring and multiplication by a constant. Let  $m_1 = X_{1}X_{2}$, $m_2 = Y_{1}Y_{2}$, $m_3 = Z_{1}Z_{2}$, $c_1=bm_{1}$ et $c_{2}=am_{2}$.
\begin{enumerate}
\item $m_{4}=(X_{1}+Z_{1})(X_{2}+Z_{2})-m_{1}-m_{3}$, \ \ $m_{5}=(Y_{1}+Z_{1})(Y_{2}+Z_{2})-m_{2}-m_{3}$
\item  $m_{6}=(m_3 - c_1)(m_3 + c_2)$, \ \  $m_{7}= (m_3 + c_1)(m_3 - c_2)$.
\end{enumerate}
Therefore,  $X_{3}= m_4 m_6(m_3 + c_2)$, $Y_{3}= m_5 m_7(m_3 + c_1)$ et $Z_{3}=m_{6}m_{7}$. Thus, the addition of two points of the curve can be evaluated in   $12\mathbf{m}+2\mathbf{c}$, where  $2\mathbf{c}$ represent the cost of the multiplications by the two constants $a$ and $b$.\\~~\\
These addition formul\ae \ are complete. In other word, they can be used to compute the point  $2P = (x_3, y_3)$ for a given point  $P = (x_1, y_1)$. Thus,
$$x_3 = \frac{2x_1(ay_1^2 + 1)}{(bx_1^2 + 1)(ay_1^2 - 1)} \ \ \text{ and } \ \ y_3 = \frac{2y_1(bx_1^2 + 1)}{(bx_1^2 -1)(ay_1^2 + 1)}.$$
In projective coordinates, the point $2$ can be evaluated in $7\mathrm{m} + 5\mathrm{s} + 2\mathrm{c}$ when working with  $U = O= (0, 0, 1)$ as neutral element

\subsection{Backgrounds on the Tate pairing}
\begin{definition}
Let $G_1$ and $G_2$ be finite abelian groups written additively, and let $G_3$ be a multiplicatively written finite group. A cryptographic pairing is a map
\begin{equation*}
e:G_1 \times G_2 \longrightarrow G_3
\end{equation*}
that satisfies the following properties:
\begin{enumerate}
\item it is  non-degenerate, ie for all $0\neq P \in G_1$, there is a $Q \in G_2$ with $e(P,Q)\neq 1$, and  for all $0\neq Q \in G_2$, there is a $P \in G_1$ with $e(P,Q)\neq 1$
\item it is bilinear, ie for all $P_1, P_2\in G_1$ and for all $Q_1, Q_2\in G_2$ we have
\begin{equation*}
e(P_1+P_2, Q_1) = e(P_1, Q_1)e(P_2, Q_1)
\end{equation*}
\begin{equation*}
e(P_1, Q_1+Q_2) = e(P_1, Q_1)e(P_1, Q_2)
\end{equation*}
\item it is efficiently computable
\end{enumerate}
\end{definition}
An important property that is used in most applications and that follows immediately from the bilinearity is $e([a]P, [b]Q)=e(P,Q)^{ab}=e([b]P, [a]Q)$ for all $a, b\in \mathbb{Z}$ and for all $(P, Q)\in G_1\times G_2$.\\~~\\
The Tate pairing can be defined on an ordinary abelian variety. It induces a pairing on the $r$-torsion subgroup of the abelian variety for a prime order $r$.\\
Let $E$ be an elliptic curve defined over a finite field $\mathbb{F}_q$ of characteristic $p$.  Let $n = \# E$ and $r > 5 $ be a prime different from $p$ and $r|n$.
\begin{definition}
The smallest integer $k$ with $r|(q^k -1)$ is called the embedding degree of $E$ with respect to $r$
\end{definition}

\begin{remark}
If $k$ is the smallest integer with $r|(q^k -1)$, the order of $q$ modulo $r$ is $k$. Furthermore, the smallest field extension of $\mathbb{F}_q$ that contains the group $\mu _r$ of all $r$-th roots of unity is $\mathbb{F}_{q^k}$.
\end{remark}
Let $E$ be an elliptic curve over $\mathbb{F}_q$ of characteristic $p > 3$ given by a short Weierstrass equation
\begin{equation*}
E: y^2 = x^3 +ax + b \ \ \, a, b\in \mathbb{F}_q.
\end{equation*}
Let $r\neq p$ be a prime such that $r|n = \#E(\mathbb{F}_q)$ and let $k > 1$ be the embedding degree of $E$ with respect to $r$.

\begin{lemma}\label{div}
Let $D = \displaystyle\sum_{P\in E} n_P(P) \in Div(E)$. Then $D$ is a principal divisor if and only if $deg(D) = 0$ and $\displaystyle\sum_{P\in E} [n_P](P) = 0$, where the latter sum describes the addition on $E$.
\end{lemma}

\begin{definition}
Let $E$ be an elliptic curve over a finite field $\mathbb{F}_q$ of characteristic $p$ and let $r\neq p $ be a prime dividing $n = \# E(\mathbb{F})$. Let $k$ be the embedding degree of $E$ with respect to $r$. The reduced Tate pairing is a map
\begin{align*}
e_r: E(\mathbb{F}_{q})[r] \times  E(\mathbb{F}_{q^k}) [r] & \longrightarrow \mu _r \subseteq \mathbb{F}_{q^k} \\
(P,Q)&\longmapsto  f_{r,P}(D_Q)^{(q^k - 1)/r}
\end{align*}
where $P\in E(\mathbb{F}_{q})[r]$ is   $\mathbb{F}_{q}$-rational point of order dividing $r$ represented by a divisor $D_p$, and  $Q\in E(\mathbb{F}_{q^k})[r]$ is $\mathbb{F}_{q^k}$-rational point  represented by a divisor $D_Q$ such that its support is disjoint from the support of $D_P$, and  $f_{r,P}\in \overline{\mathbb{F}}_{q^k}(E)$ is a function on $E$ with $\mathrm{div}(f_{r,P})=rD_P$.
\end{definition}
When computing $f_{r,P}(Q)$ , ie when $rD_P$ is supposed to be the divisor of the function $f_{r, P}$, we can choose $D_P = (P)-(O)$. The divisor $D_Q \sim (Q)-(O)$ needs to have a support disjoint from $\{O, P\}$. To achieve that, one may choose a suitable point $S\in E(\mathbb{F}_{q^k})$ and represent $D_Q$ as $(Q+S)-S$.\\
We need to compute $f_{r, P}$ having divisor $\mathrm{div}(f_{r,P}) = r(P)- r(O)$. Lemma \ref{div} shows that for $m\in \mathbb{Z}$, the divisor $m(P)-([m]P) - (m-1)(O)$ is principal, such that there exists a function $f_{m,P}\in \overline{\mathbb{F}}_q(E)$ with $\mathrm{div}(f_{m,P})=m(P)-([m]P) - (m-1)(O)$. Since $P$ is a $r$-torsion point, we see that $\mathrm{div}(f_{r,P}) = r(P)-r(O)$, and $f_{r,P}$ is a function we are looking for.

\begin{definition}
Given $m\in \mathbb{Z}$ and $P\in E(\mathbb{F}_{q^k})[r]$, a function $f_{m,P}\in \overline{\mathbb{F}}_{q^k}(E)$ with divisor $\mathrm{div}(f_{m,P}) = m(P)-([m]P) - (m-1)(O)$ is called a Miller function
\end{definition}

\begin{lemma}
Let $P_1, P_2\in E$. Let $l_{P_1, P_2}$ be the homogeneous polynomial defining the line through $P_1$ and $P_2$, being the tangent to the curve if $P_1 = P_2$. The function $L_{P_1, P_2} = l_{P_1, P_2}(X,Y,Z)/Z$ has the divisor
\begin{equation*}
\mathrm{div}(L_{P_1, P_2}) = (P_1) + (P_2) + (-(P_1 + P_2)) - 3(O).
\end{equation*}
\end{lemma}

\begin{lemma}
Let $P_1=(x_1, y_1)$, $P_2 = (x_2, y_2)$, $Q = (x_Q, y_Q)\in E$. For $P_1\neq -P_2$ define
\begin{equation*}
    \lambda=\left\{\begin{array}{rl}
						(y_2-y_1)/(x_2-x_1) & \quad\mbox{if }  P_1\neq P_2, \\
						(3x_1^2+a)/(2y_1)\ \ \quad & \quad\mbox{if }  P_1=P_2.\end{array}\right.
\end{equation*}
Then, the dehomogenization  $(l_{P_1, P_2})_{*}$ of $l_{P_1, P_2}$ evaluated at $Q$ is given by
\begin{equation*}
(l_{P_1, P_2})_*(Q) = \lambda(x_Q - x_1) + (y_1 - y_Q).
\end{equation*}
If $P_1 = - P_2$, then $(l_{P_1, P_2})_*(Q) =x_Q - x_1$.
\end{lemma}

\begin{lemma}
Let $P_1, P_2\in E$. The function $g_{P_1, P_2}:= L_{P_1, P_2}/L_{P_1 + P_2, -(P_1+P_2)}$ has the divisor
\begin{equation*}
\mathrm{div}(g_{P_1, P_2}) = (P_1) + (P_2) - (P_1 + P_2) -(O).
\end{equation*}
\end{lemma}
The function $g$ can be used to compute the Miller function recursively as shown in the next lemma.

\begin{lemma}
The Miller function $f_{r,P}$ can be chosen such that $f_{1, P} = 1$  and such that for $m_1, m_2 \in \mathbb{Z}$, it holds
\begin{equation*}
f_{m_1+m_2,P} = f_{m_1, P}f_{m_2, P}g_{[m_1]P, [m_2]P},
\end{equation*}
\begin{equation*}
f_{m_1 m_2,P} = f_{m_1, P}^{m_2}f_{m_2, [m_1]P}= f_{m_2, P}^{m_1}f_{m_1, [m_2]P}
\end{equation*}
\end{lemma}

\begin{remark}
Special cases from the previous lemma\\
Let $m\in \mathbb{Z}$, then
\begin{enumerate}
\item $f_{m+1, P} = f_{m,P}g_{[m]P, P}$,
\item $f_{2m, P} = f_{m,P}^2 g_{[m]P, [m]P}$,
\item $f_{-m, P} = (f_{m,P}g_{[m]P}, -[m]P)^{-1}$.
\end{enumerate}
Note that $f_{0, P} = 1$ for all $P\in E$ and $g_{P_1, P_2} = 1$ if $P_1$  or $P_2$ equals the point at infinity $O$. These formulas show that any function $f_{m,P}$ can be computed recursively as a product line functions. The functions are defined over the field of definition of $P$.
\end{remark}

\begin{lemma}
Let $P\in E(\mathbb{F}_q)[r]$ and $Q\in E(\mathbb{F}_{q^k})[r]$, $\notin E(\mathbb{F}_q)$, then the reduced Tate pairing can be computed as $e_{r}(P,Q) = f_{r, P}(Q)^{(q^k - 1)/r}$.
\end{lemma}
The above algorithm, well known as the Miller's algorithm, can be used to compute $f_{r,P}(Q)$ for $P\in E(\mathbb{F}_q)[r]$ and $Q\in E(\mathbb{F}_{q^k})[r]$ and $r = (r_l, r_{l-1},\ldots, r_0 )_2$ up to irrelevant factors lying a proper subfield of $\mathbb{F}_{q^k}$. Since $k > 1$, these factors are mapped to 1 by the final exponentiation.

\begin{algorithm}[t]
\caption{Miller's Algorithm}
\label{miller_algo}
\begin{algorithmic}[1]
\STATE $R\gets P$, $f\gets 1$
\FOR {($i=l-1$; $i \geq$ 0; $i--$)}
   \STATE  $f\gets f^2.g_{R,R}(Q)$
   \STATE  $R\gets 2R$
   \IF {($r_i = 1$)}
   			\STATE $f \gets f.g_{R,P}(Q)$
   			\STATE $R \gets R+P$
   \ENDIF
\ENDFOR
\RETURN $f^{(q^k - 1)/r}$
\end{algorithmic}
\end{algorithm}

\begin{remark}
Note that the functions $g_{R,R}$ and $g_{R,P}$ in steps 3 and  6 are fractions and that the inversions in each step of the loop can be postponed until the end of the loop by keeping track of numerator and denominator separately.
\end{remark}

\section{Pairing computation on generalized Huff curves}
The Tate pairing computation on the classic Huff curves  was introduced by Joye \emph{et al.} in \cite{Joye}. The main contribution of this paper is the extension of the previous results on the generalized Huff curves.\\

Huff curves can be represented as plane cubics. Thus, we can apply directly the Miller Algorithm to compute pairings on these curves. It's quite usual to represent the point  $Q\in E(\mathbb{F}_{q^k}) \backslash E(\mathbb{F}_{q})$ in affine coordinates since, in the Miller algorithm, the function is always evaluated at the same point. Let  $Q = (y,z) = (1:y:z)$. Suppose the embedding degree  $k$ is even, then  $Q$ can be written in the form   $Q=(y_Q, z_Q \alpha)$, with $y_Q, z_Q\in \mathbb{F}_{q^{k/2}}$, $\mathbb{F}_{q^{k}}=\mathbb{F}_{q^{k/2}}(\alpha)$, where $\alpha$ is a non quadratic residue in  $\mathbb{F}_{q^{k/2}}$.\\
Let   $P,R\in E(\mathbb{F}_{q})$ and  $l_{R,P}$ be the rational function vanishing on the line through  $P$ and $R$. We have
\begin{equation*}
l_{R, P}(Q) = \displaystyle\frac{(zX_p - Z_p) - \lambda(yX_p - Y_P)}{X_P}
\end{equation*}
where $\lambda$ is the $(y,z)$-slope of the line through  $P$ and $R$. Then, the divisor of $l_{R,P}$ is given by
\begin{equation*}
\mathrm{div}(l_{R,P}) = R + P + T -(1:0:0) - (0:1:0) - (a:b:0)
\end{equation*}
where is $T$ is the third intersection of the line through   $P$ and  $R$ with the curve.  If $U$ is the neutral element of the group law (+), then the function $g_{R,P}$ can be expressed as
\begin{equation*}
g_{R,P} = \displaystyle\frac{l_{R,P}}{l_{R + P, U}}
\end{equation*}
Let   $U=O=(0:0:1)$ be the neutral element of the addition law. Then, for all  $Q = (y_Q, z_Q \alpha)$, we have
\begin{equation*}
l_{R + P, O} = y_Q - \displaystyle\frac{Y_{R + P}}{X_{R + P}} \in \mathbb{F}_{q^{k/2}}
\end{equation*}
This quantity is equal to 1 after the final exponentiation in the Miller algorithm since it belongs to a proper sub-field of  $\mathbb{F}_{q^k}$. That means it can be canceled in computations. In the same context, divisions by  $X_P$ can be omitted, and the denominator in the expression of  $\lambda$ too, ie. if  $\lambda = \displaystyle \frac{A}{B}$, then the function  $g_{R,P}$ can be evaluated as
\begin{equation*}
g_{R,P}(Q) = (z_Q \alpha .X_p - Z_p)B - (yX_p - Y_P)A
\end{equation*}
We are now ready to  give explicit and precise formul\ae \ for the addition and doubling steps of each round of the Miller loop.\\~~\\
\textbf{Addition step}. In the addition step, the $(y,z)$-slope of line through the points $P = (X_P : Y_P : Z_P)$ and  $R = (X_R : Y_R : Z_R)$ is given by
$$\lambda = \frac{Z_R X_P - Z_P X_R}{Y_R X_P - Y_P X_R}.$$
Therefore, the function to be evaluated is of the form
$$g_{R,P}(Q) = (z_Q \alpha . X_P - Z_P)(Y_R X_P - Y_P X_R) - (y_Q . X_P - Y_P)(Z_R X_P - Z_P X_R).$$
Since the points  $P$ and $Q$ remain constant during the execution of the Miller loop, the values depending on  $P$ and $Q$, ie.  $y_Q' = y_Q.X_P - Y_P$ and  $z_Q' = z_Q\alpha . X_P$ can be precomputed. Thus, each addition step of the Miller algorithm requires the calculation of  $R +  P$ (an addition over  $E(\mathbb{F}_q)$), the evaluation of $g_{R,P}(Q)$, and the calculation of  $f.g_{R,P}(Q)$ (a multiplication over the field extension $\mathbb{F}_{q^k}$).\\
$R + P$ can be evaluated in   $12 \mathrm{m} + 2 \mathrm{c}$ using the steps  $M_1, M_2, \ldots, M_7$.\\Let   $m_{8} = (X_R + Y_R)(X_P - Y_P)$ and  $m_{9} = (X_P + Z_P)(Z_R - X_R)$. Then,
\begin{equation*}
g_{R,P}(Q) = (z_Q ' - Z_P)(m_{8} - m_1 + m_2) - y_Q'(m_{9} + m_1 - m_3) ,
\end{equation*}
where the first term require  $(\frac{k}{2} + 1) \mathbf{m}$, and the second one  $\frac{k}{2} \mathbf{m}$. With the final multiplication in  $\mathbb{F}_{q^k}$, the cost of an addition step is  $1\mathbf{M}+(k+15)\mathbf{m}+2\mathbf{c}$.\\
~~\\
\textbf{Doubling step.} In the doubling step, the slope of the tangent  line to the curve at the point $R =(X_R : Y_R : Z_R)$ is given by $$\lambda = \frac{aZ_R^2 -2bY_R Z_R - X_R^2}{bY_R^2 - 2aY_R Z_R - X_R^2} = \frac{A}{B}.$$
Therefore,
$$g_{R,R}(Q) = z_Q\alpha. X_R B - Z_R B - y_Q . X_R A + Y_R A .$$
In the Miller algorithm, we need to compute the point $2R$, which can be done in  $7\mathbf{m} + 5\mathbf{s} + 2\mathbf{c}$.  $A$ and $B$ can be evaluated in  $1\mathbf{m}$, namely $Y_R Z_R$ since the other terms are already computed with the doubling operation. Therefore, the function  $g_{R,R}$ can be computed in  $4\mathbf{m}$ ($X_R B$, $Z_R B$, $X_R A$ and $Y_R A$), $\frac{k}{2}\mathbf{m}$ for $z_Q\alpha . X_R B $ and $\frac{k}{2}\mathbf{m}$ for $y_Q . X_R A $. Thus, the doubling step require a total cost of  $1\mathbf{M} + 1\mathbf{S} + (k + 12) \mathbf{m} + 5\mathbf{s} + 2\mathbf{c}$, by taking in account the multiplication, the squaring which complete the duplication.

\section*{Conclusion}	We have successfully extended the Tate pairing computation on generalized Huff curves introduced by Wu and Feng. Our results are not far from the standard case since the the multiplication by constant are often negligible. That makes it as efficient as the standard Tate pairing computation on Huff curves proposed by  Joye, Tibouchi and Vergnaud. The next step is to use use result to design efficient cryptographic protocols such as ID-base protocols.

\end{document}